# Formation of Random Dark Envelope Solitons from Incoherent Waves

Wei Tong,[1] Mingzhong Wu,[1]* Lincoln D. Carr,[2] and Boris A. Kalinikos[3]

[1]*Department of Physics, Colorado State University, Fort Collins, Colorado 80523, USA*

[2] *Department of Physics, Colorado School of Mines, Golden, Colorado 80401, USA*

[3]*St.Petersburg Electrotechnical University, 197376, St. Petersburg, Russia*

**Corresponding author. E-mail: mwu@lamar.colostate.edu*

This letter reports experimental results on a new type of soliton: the random temporal dark soliton. One excites an incoherent large-amplitude propagating spin-wave packet in a ferromagnetic film strip with a repulsive, instantaneous nonlinearity. One then observes the random formation of dark solitons from this wave packet. The solitons appear randomly in time and in position relative to the entire wave packet. They can be gray or black. For wide and/or very strong spin-wave packets, one also observes multiple dark solitons. In spite of the randomness of the initial wave packets and the random formation processes, the solitons show signatures that are found for conventional coherent dark solitons.

Solitons are a ubiquitous phenomenon in nature, appearing in systems as diverse as water, optical fibers, deoxyribonucleic acid (DNA), and ultra-cold atoms [1-5]. Such solitons are described by well-known paradigmatic nonlinear equations, one of which is the nonlinear Schrödinger equation (NLSE). In the terms of the NLSE model, two classes of envelope solitons, bright and dark, can exist in nonlinear media. Bright envelope solitons are localized large-amplitude excitations on the envelope of certain carrier waves. Their formation requires an attractive or focusing nonlinearity. Dark envelope solitons are dips or holes in a large-amplitude wave background. Their formation requires a repulsive or defocusing nonlinearity.

Two classes of dark solitons, temporal and spatial, can propagate in systems with a repulsive nonlinearity [6-9]. A temporal dark soliton is a dip on a temporal continuous wave. When the dip goes to zero, one has a black soliton. When the amplitude at the dip is nonzero, one has a gray soliton. A spatial dark soliton is a low-intensity hole in a high-intensity background. Like temporal dark solitons, spatial dark solitons can also be black or gray, which depends on the intensity at the soliton regions. Both temporal and spatial dark solitons have a jump in phase at their centers. For gray solitons, such a phase jump is between 0° and 180°. For black solitons, the phase jump equals exactly 180°. The envelope of temporal dark solitons and the cross-section profile of spatial dark solitons can be described by one and the same function. For black solitons, this function is a hyperbolic tangent function.

Like bright solitons, for many years dark solitons have been taken as *coherent* nonlinear entities developed from *coherent* waves. However, about ten years ago numerical and experimental work demonstrated that dark solitons can also be formed from partially *incoherent* waves [10-14]. This was realized for the case of *spatial* optical dark solitons. In comparison with conventional dark solitons, such dark solitons have several unique properties. First, they are incoherent or have random phase. Second, the solitons are always gray and cannot be black. Third, the media must have a *noninstantaneous* nonlinearity, i.e., a nonlinear response time that is much longer than the fluctuation or correlation time of the incoherent wave.

It is important to emphasize that previous work on incoherent dark solitons was for the case of spatial solitons and noninstantaneous nonlinearity. Can a temporal dark soliton be formed from incoherent waves? If yes, what are the properties of such a soliton? Do the media need to have a noninstantaneous nonlinearity? Is it possible to have a fundamental black soliton? These questions are important and intriguing but have never been addressed before.

This letter reports the random formation of *temporal* dark solitons from incoherent waves that have an *instantaneous* nonlinearity, i.e., a nonlinear response time much shorter than the correlation time of the incoherent waves. One excites a packet of large-amplitude incoherent waves in a one-dimensional medium with a repulsive nonlinearity. One then follows the propagation of such a wave packet and observes the formation of dark solitons from the wave packet. There are three important features of these solitons. (1) The solitons have a random nature. They appear randomly in time and in position relative to the overall wave packet. (2) The solitons can be both gray and black. (3) In spite of the random nature of the initial wave packet and the random formation process, the solitons exhibit the same kind of properties as conventional dark solitons. More specifically, they are coherent and have the same shape and phase properties as conventional dark solitons. Therefore, the solitons described in this letter are fundamentally a new concept. They differ significantly from both conventional dark solitons [6-9] and incoherent dark solitons [10-14].

The experiment was performed for surface spin waves in a magnetic yttrium iron garnet (YIG) film strip. Surface spin waves have a repulsive nonlinearity, and the formation of dark solitons from coherent surface spin waves has been demonstrated previously [7, 15-17]. Pulses of noisy microwaves were used to excite partially incoherent spin-wave packets at one end of the YIG strip. The propagation of such packets along the YIG strip was probed by microstrip line transducers [18]. The nonlinear response time of the YIG film is inversely proportional to the power of the spin waves [18, 19]. The correlation time of the incoherent packet is determined by the bandwidth of the incoherent spin-wave signals. When the spin-wave amplitude is sufficiently large to push the nonlinear response time below the correlation time, the spin waves experience an instantaneous nonlinearity and random dark solitons appear. It is worth noting that the excitation of pulsed spin waves, rather than continuous spin waves, allowed for the easy tracing of a specific signal at different positions along the propagation path. However, the results reported below are also applicable to continuous waves.

Figure 1 shows the experimental setup. The magnetic field is in the plane and perpendicular to the length of the YIG strip. This arrangement supports the propagation of surface spin waves [20,21]. Three microstrip line transducers are positioned over the YIG strip. The left-most one is for the excitation of spin waves, and the others are for the detection of spin waves.

For the experimental data presented below, the YIG strip was 7.2 μm thick, 2 mm wide, and 50 mm long. It was cut from a larger single-crystal YIG film grown on a gadolinium gallium garnet substrate by liquid phase epitaxy [22]. The spins on the surfaces of the YIG film were unpinned. The spin waves in such a film can be described by a simple monotonic dispersion curve [20]. The magnetic field was 1113 Oe. The microstrip line transducers were 50 μm wide and 2 mm long. The detection transducers were held at displacements ($x$) of 7 mm and 15 mm from the input transducer. The band of microwave noise ranged from about 2 GHz to about 8 GHz. The frequency band of available spin-wave modes ranged from about 5.0 GHz to about 5.4 GHz. The noise band was much wider than the spin-wave band and therefore allowed for the excitation of all available spin-wave modes. A statistical analysis confirmed the random nature of the initial noisy signals. The two-point autocorrelation function was a near-δ function with a half-power width of about 0.32 ns and a peak-to-background ratio of about 450. The 3-point and 4-point autocorrelations also took the form of near-δ functions.

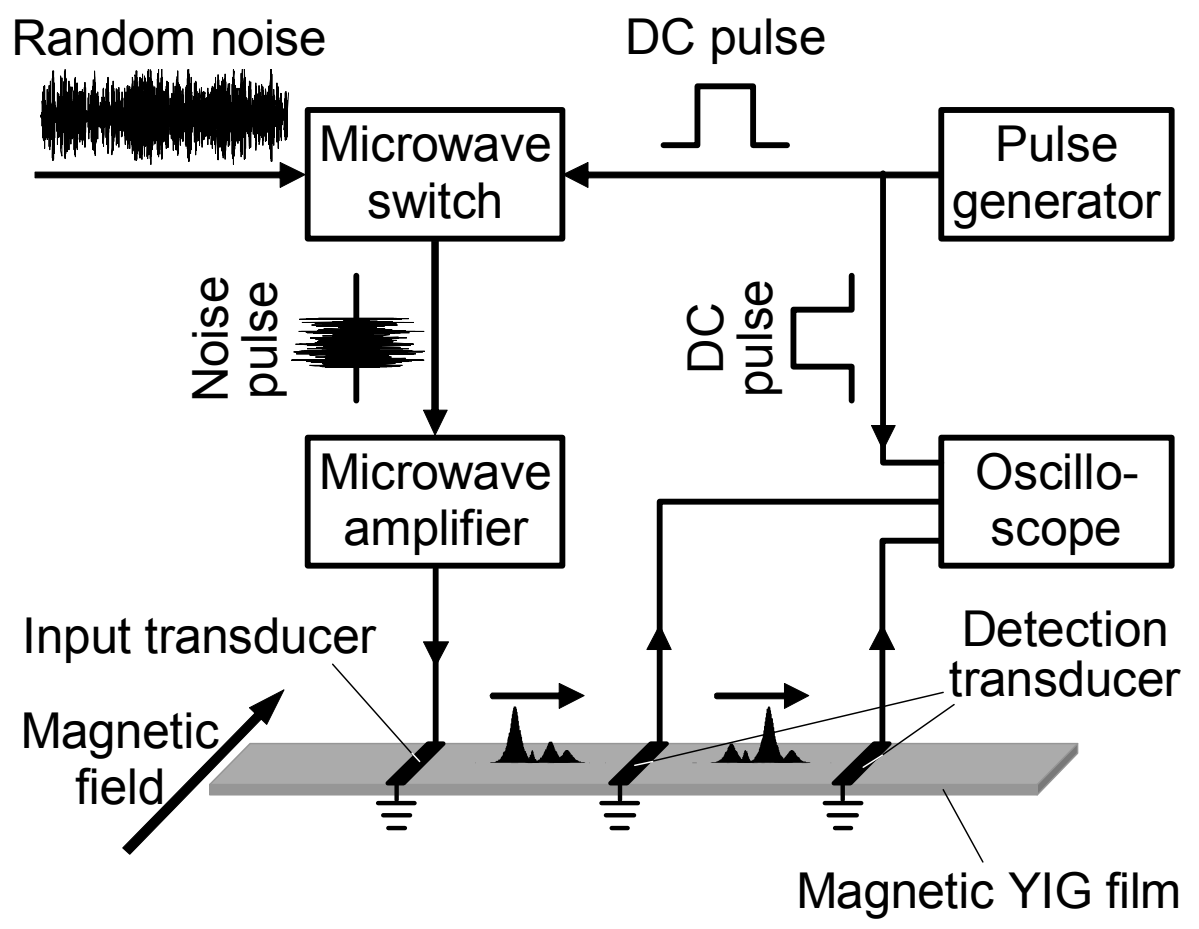

FIG. 1. Schematic of the experimental arrangement.

The waveforms shown below were recorded directly by a broadband oscilloscope. The phase profile for a given dip is the phase of the carrier wave relative to a reference continuous wave signal. The frequency of the reference signal was given by the main frequency of the carrier wave of the soliton. This main frequency was determined through a fast Fourier transform analysis. Note that the relative phase profiles allow for a better presentation of the phase jumps in the centers of solitons.

The amplitude and phase profiles of the solitons were fitted theoretically through the general form of a gray soliton, a solution to the NLSE as described in Ref. [8]:

$$u(t) = a_0\left\{i\sqrt{1-a_1^2} + a_1 \tanh\left[a_2(t-a_3)\right]\right\}\exp\left[i\left(a_4 t + a_5\right)\right],$$

where $\{a_0,\ldots,a_5\}$ are fit parameters, $|u|$ is the amplitude, and arg($u$) gives the phase. This fit function was applied to selected regions of the experimental data using a least-squares fitting routine. Physically, the meaning of the fit parameters is as follows: $a_0$ determines the background level of the soliton; $a_1$ determines the depth or "grayness" of the soliton; $a_2$ determines the width and nonlinearity; $a_3$ determines the displacement from the time origin; $a_4$ determines the local slope of the phase, which is proportional to the velocity of the flat amplitude region away from the soliton; and $a_5$ determines the phase offset, an important symmetry of the nonlinear Schrödinger equation. Because the fits were instantaneous, spatial dependence was neglected in the fitting function.

Figure 2 shows representative data for the random formation of dark solitons. Graphs (a)-(c) give three sets of waveforms sampled for three input noise pulses, each with a width of 25 ns and a power of 5 dBm. In each graph, the left and right diagrams show the waveforms detected at different positions as indicated. The insert in (c) shows the waveform in an expanded horizontal scale. The arrows and labels identify the dips, which are solitons.

Several results are evident in Fig. 2. (1) For a given noise pulse, the solitons appear randomly. For example, in Sample 1 the leading dip is a gray soliton at $x$=7 mm. However, at $x$=15 mm one sees a black soliton at the center of the overall packet. In contrast, in Sample 2 one sees a black soliton at $x$=7 mm but a gray soliton at $x$=15 mm. In Sample 3, the leading dip is a gray soliton at $x$=7 mm. However, at $x$=15 mm the leading dip is not a soliton but the second dip is a black soliton. (2) The overall waveforms and the appearance of gray and black solitons are random from sample to sample. (3) Both gray and black solitons are possible. Note that previous work reported that, in the case of incoherent spatial dark solitons, the formation of black solitons was impossible [10, 12]. (4) For each soliton, the carrier wave is coherent as shown representatively by the insert in (c). This result is further confirmed by the coherent phase profiles shown in Fig. 3. (5) The overall wave packet at *x*=15 mm is much wider than that at *x*=7 mm, and both are much wider than the initial wave packet. This indicates that the initial incoherent spin-wave packet disperses significantly during its propagation along the YIG strip.

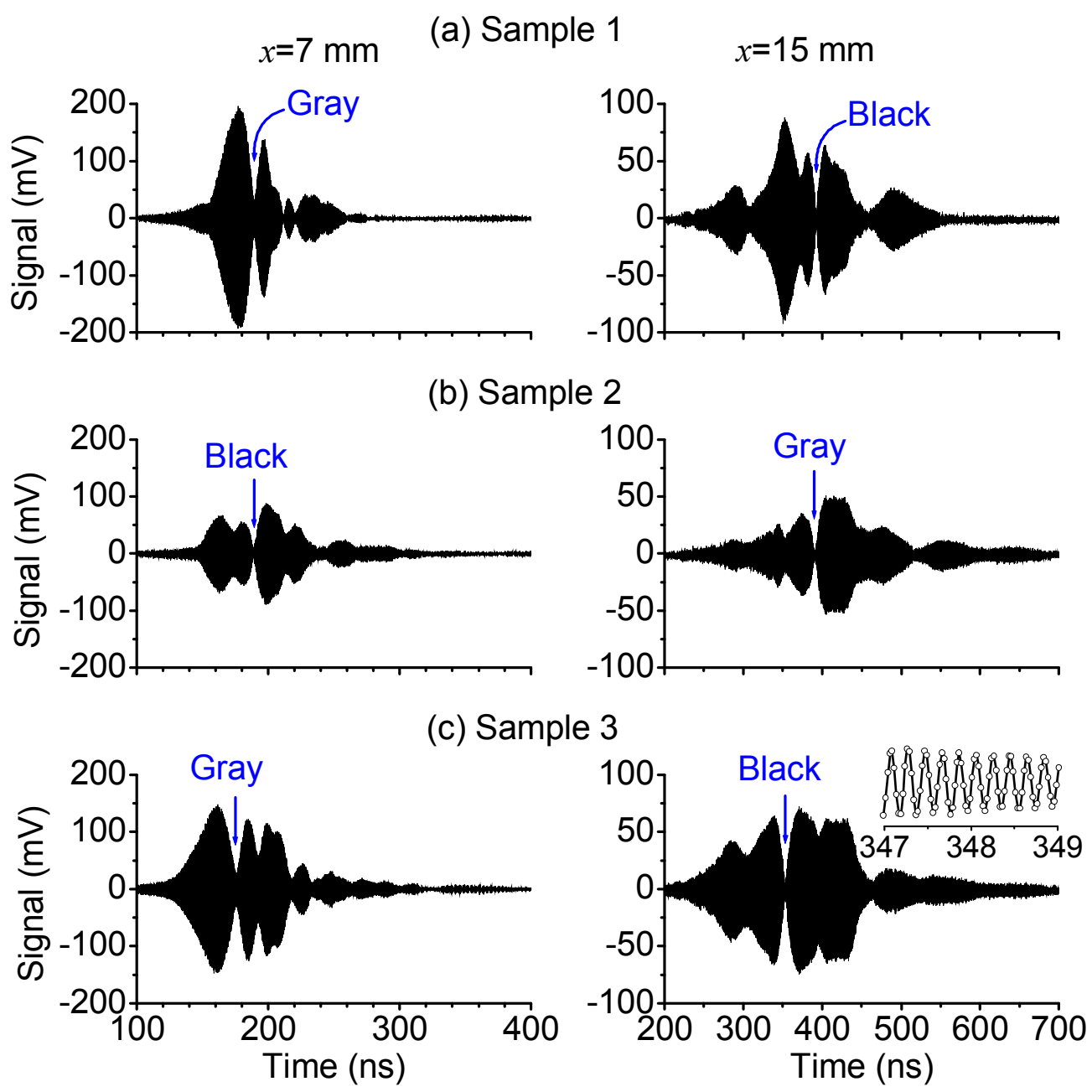

FIG. 2. Three sets of waveforms sampled for three input noise pulses, each with a width of 25 ns and a power of 5 dBm. Left and right diagrams show the waveforms detected at different positions, as indicated. The insert in (c) shows the waveform in an expanded horizontal scale. The arrows and labels identify the dips which are solitons.

Figure 3 presents the characteristics for the two solitons shown in Fig. 2(a). Graphs (a) and (b) are for the solitons in the left and right diagrams in Fig. 2(a), respectively. In each graph, the top and bottom diagrams show the amplitude and phase profiles, respectively. The circles show experimental data. The curves show theoretical fits. The vertical dashed lines indicate the center of solitons.

The data in Fig. 3 clearly indicate that the observed solitons have the signatures found for conventional coherent dark solitons. First, one sees almost perfect matches between the experimental and theoretical soliton profiles. This indicates that the solitons have the same shape as conventional dark solitons. In particular, the black soliton has a zero dip and a hyperbolic tangent shape. Second, both solitons have a phase jump at their centers. For the gray soliton, this phase jump is about 90°, which is less than 180° as usual. For the black soliton, however, one sees a phase change of about 180°. Third, the phase profiles are smooth and clean, and their fits are self-consistent with the amplitude profiles. This supports the conclusion that the solitons are coherent. It is important to emphasize that, for each soliton, the amplitude and phase profiles were fitted with the same set of parameters.

The data in Figs. 2 and 3 are for a relatively narrow and low-power noise input pulse. With an increase either in pulse width or in power, one observes random dark soliton pairs and triplets. Figure 4 show representative data. Column (a) shows the data for a noise input pulse that has the same width as the pulses for the above-discussed data but a much higher power level of 25 dBm. Column (b) shows the data for the 100 ns and 20 dBm input pulse. Both data were measured at $x$=7 mm. In each column, the top diagram shows the waveform. The middle and bottom diagrams show the amplitude and phase profiles, respectively, for the solitons marked with a dashed outline in the top diagram. In the middle and bottom diagrams, the circles show experimental data, the curves show theoretical fits, and the vertical dashed lines indicate the centers of the solitons.

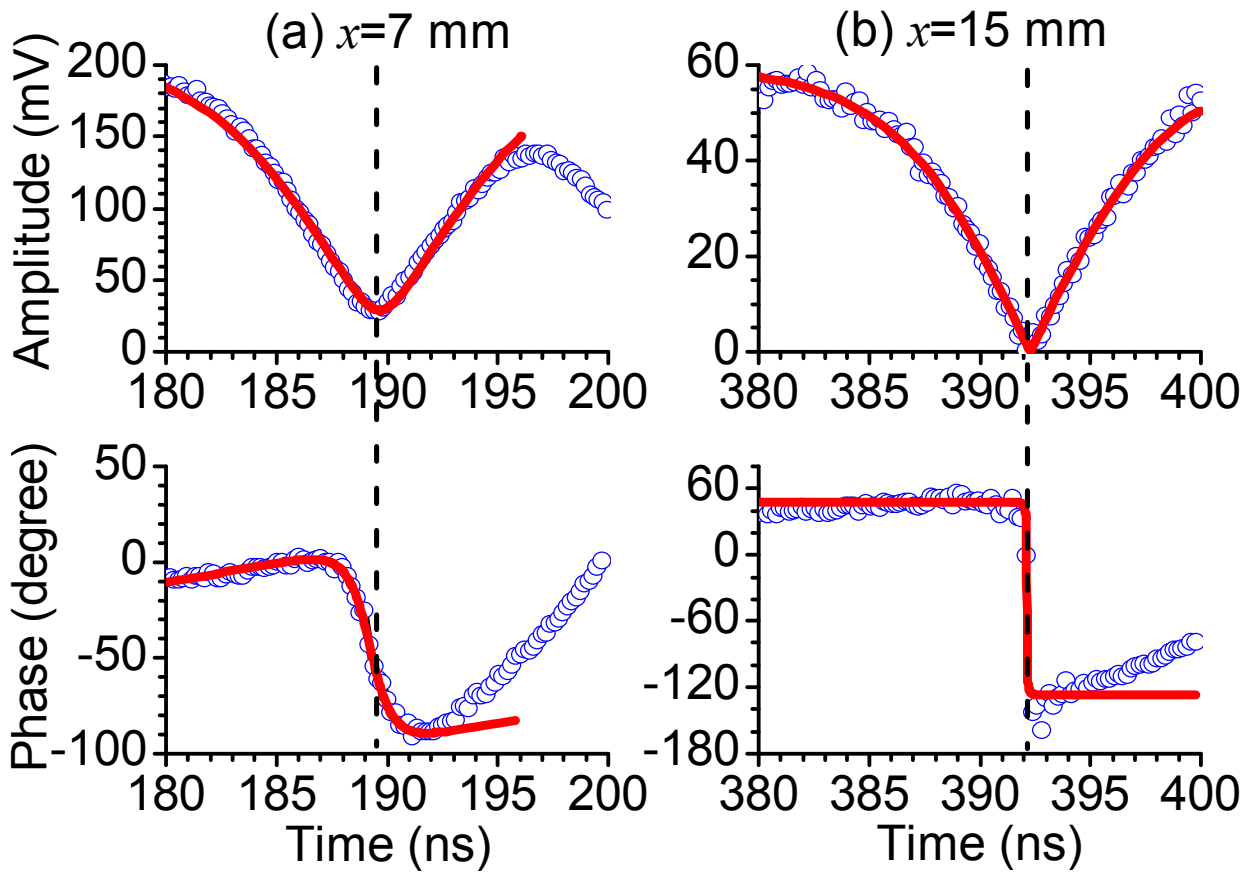

FIG. 3. Characteristics of dark solitons shown in Fig. 2(a). Graphs (a) and (b) are for the signals detected at different positions, as indicated. In each graph, the top and bottom diagrams show the amplitude and phase profiles, respectively. Circles are experimental data. Curves are theoretical fits.

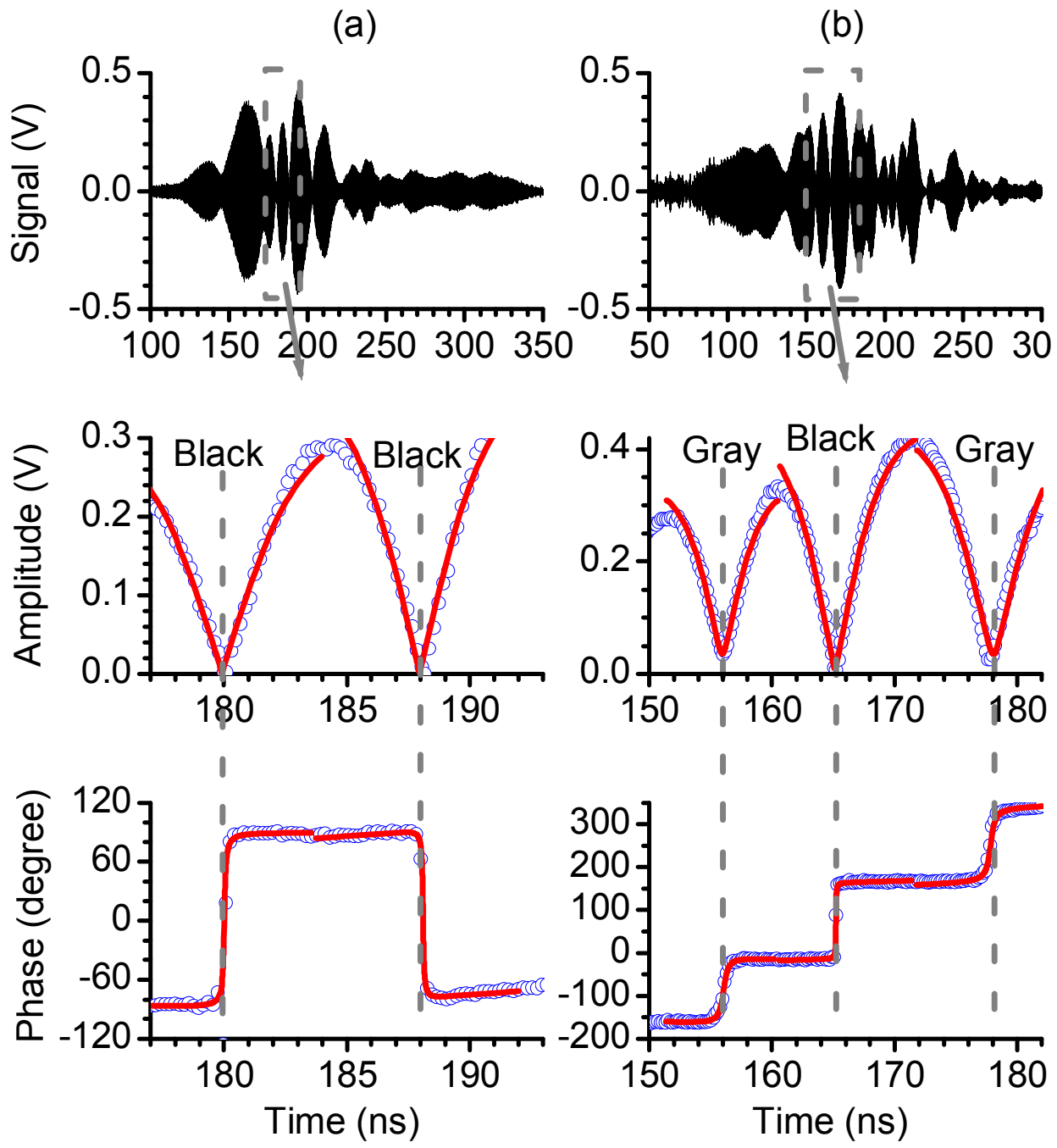

FIG. 4. Data and fits for (a) a 25 ns and 25 dBm input pulse and (b) a 100 ns and 20 dBm input pulse. In each column, the top, middle, bottom diagrams show the waveform, amplitude profile, and phase profile, respectively. Circles are experimental data. Curves are theoretical fits.

The data in Fig. 4(a) show a black "soliton pair." Both solitons have a zero dip, a hyperbolic tangent shape, and a smooth phase profile with a 180° phase jump at the dips. The data in Fig. 4(b) show a completely different response. Specifically, the data show a gray-black-gray "soliton triplet." It is important to emphasize that, although not shown here, the multiple solitons appear randomly from position to position and randomly from sample to sample.

It is worth mentioning that the direction of soliton phase changes is essentially random. For the single solitons in Fig. 2, one observes both up-jump and down-jump phase changes. For the soliton pair in Fig. 4(a), the phase changes are opposite; one phase jumps up and the other jumps down. For the solitons in Fig. 4(b), in contrast, the phases all jump up and one sees a step response.

The data in Figs. 2-4 clearly demonstrate the formation of random temporal dark solitons from incoherent waves. Here, "random" means three things. First, the solitons appear randomly in time and in position relative to the entire wave packet. Second, the type of soliton is random. They can be gray or black and single or multiple in number. Third, the direction of the phase change is random. A statistical analysis of fit parameters over 121 mainly black solitons found the parameters random over the approximate ranges $a_0 \in [0,0.4]$, $a_2 \in [-0.5,0.5]$, $a_3$ an equivalent spread around two measured places on the film, $a_4 \in [-0.05,0.05]$, and $a_5 \in [0,2\pi)$. In spite of their random nature, the solitons show signatures found for conventional dark solitons.

The essential physical process for the formation of the above-described random solitons is similar to the previous work on random bright solitons [23]. The main mechanisms underlying this process are as follows. First, the propagating spin-wave packets consist of a large number of uncorrelated spin-wave modes. Second, during the propagation of the packet, the spin-wave modes experience different types of interferences that lead to the random peaks and dips in the wave packet. Third, a dip that has a strong background experiences an instantaneous nonlinearity and thereby can evolve into a dark soliton quickly. Once realized, the soliton behaves like a conventional dark soliton over a time interval that is shorter than the correlation time. The soliton loses its soliton features beyond the correlation time.

For the situation depicted in Figs. 2 and 3, the nonlinear response time of the YIG film and the correlation time of the incoherent spin-wave packet were estimated to be about 3 ns and 18 ns, respectively. The details on the estimation of these times were given in Refs. [18, 19, 23]. These time scales clearly indicate that the media have an instantaneous nonlinearity. It is this instantaneous nonlinearity that facilitates the formation of dark solitons discussed above. The random nature of the soliton formation results from the incoherent nature of the spin-wave packets. The phase of the initial waves changes randomly, and this leads to the randomness of the direction of phase changes.

There are two additional points of note. First, the random solitons presented above differ significantly from coherent spin-wave dark solitons reported in Refs. [7, 15-17]: (1) the initial waves were incoherent waves (i.e., many modes with random phase); (2) the formation and types of the solitons were random; and (3) the solitons had a life time much shorter than the coherent solitons. Second, the analysis indicates that an effective short-time-scale description by a nonlinear Schrödinger equation with random parameters is an effective model. This presents a significant challenge for future theoretical work.

The authors acknowledge Carl Patton and John Scales for useful discussions. This work was supported in part by the U. S. Army Research Office, the U. S. National Science Foundation, and the Russian Foundation for Basic

Research.